\documentclass[twocolumn]{aastex631}
\usepackage{graphicx,natbib,color,bm,url,times}
\usepackage{amsmath}
\usepackage{amssymb}
\usepackage{euscript}
\usepackage{hyperref}

\newcommand{\beq}{\begin{equation}}
\newcommand{\eeq}{\end{equation}}

\newcommand{\bfu}{\mbox{\boldmath $u$}}

\newcommand{\bfB}{\mbox{\boldmath $B$}}

\newcommand{\rem}{{\rm Rm}}
\newcommand{\re}{{\rm Re}}

\newcommand{\sh}{{\rm Sh}}

\newcommand{\ma}{{\rm Ma}}

\def\half{{\textstyle{1\over2}}}
\newcommand{\JJ}{\mbox{\boldmath $J$} {}}
\newcommand{\const}{{\rm const}  {}}
\newcommand{\csu}{c_{\rm su}}
\newcommand{\csd}{c_{\rm sd}}
\newcommand{\cs}{c_{\rm s}}
\newcommand{\urms}{u_{\rm rms}}
\newcommand{\Brms}{B_{\rm rms}}
\newcommand{\Bym}{\overline{B}_{y}}
\newcommand{\Bxm}{\overline{B}_{x}}
\def\cp{c_{\rm p}}
\def\cv{c_{\rm v}}

\newcommand{\Fig}[1]{Figure~\ref{#1}}
\newcommand{\Figs}[2]{Figures~\ref{#1} and \ref{#2}}
\newcommand{\EQ}{\begin{equation}}
\newcommand{\EN}{\end{equation}}
\newcommand{\EQA}{\begin{eqnarray}}
\newcommand{\ENA}{\end{eqnarray}}

\newcommand{\Eq}[1]{Eq.~(\ref{#1})}

\newcommand{\App}[1]{Appendix~\ref{#1}}
\newcommand{\Sec}[1]{Sect.~\ref{#1}}

\newcommand{\Tab}[1]{Table~\ref{#1}}

\shorttitle{Magnetic field reversals in accretion disc}
\shortauthors{Singh et al.}

\begin{document}

\title{Reversals of toroidal magnetic field in local shearing box simulations of
accretion disc with a hot corona}
\correspondingauthor{Nishant K. Singh}
\email{Email: nishant@iucaa.in, arunima.ajay@sdcollege.in,
srr@sdcollege.in}

\author[0000-0001-6097-688X]{Nishant K. Singh}
\affiliation{
Inter-University Centre for Astronomy \& Astrophysics, Post Bag 4, Ganeshkhind, Pune 411 007, India
}

\author{Arunima Ajay}
\affiliation{
Department of Physics and Research Centre, S D College, University of Kerala, India
}
\affiliation{
Research Centre, University of Kerala, India
}
\author[0000-0002-7717-0521]{S  R Rajesh}
\affiliation{
Department of Physics and Research Centre, S D College, University of Kerala, India
}



\begin{abstract}
Presence of a hot corona above the accretion disc can have
important consequences for the evolution of magnetic fields and the Shakura-Sunyaev (SS) viscosity parameter $\alpha$ in such a
strongly coupled system.
In this work, we have performed
three-dimensional magnetohydrodynamical (3D-MHD) shearing-box numerical
simulations of accretion disc with a hot corona above the cool disc.
Such a two-layer, piece-wise isothermal system is vertically stratified under linear gravity and initial
conditions here include a strong azimuthal magnetic ﬁeld
with a ratio between the thermal and magnetic pressures being of
order unity in the disc region. Instabilities in this magnetized system
lead to the generation of turbulence, which, in turn, governs the
further evolution of magnetic fields in a self-sustaining manner.
Remarkably, the mean toroidal magnetic field undergoes a
complete reversal in time by changing its sign, and it is predominantly
confined within the disc. This is a rather unique class of evolution
of the magnetic field which has not been reported earlier. Solutions
of mean magnetic fields here are thus qualitatively different from
the vertically migrating dynamo waves that are commonly seen in
previous works which model a single layer of an isothermal gas. 
Effective $\alpha$ is found to have values between
0.01 and 0.03. We have also made a comparison between models with
Smagorinsky and explicit schemes for the kinematic viscosity ($\nu$).
In some cases with an explicit $\nu$ we find a
burst-like temporal behavior in $\alpha$.
\end{abstract}

\section{Introduction}
A wide range of regimes of astrophysical MHD find their application in accretion discs around compact objects, which make these systems among the most explored in astrophysics for the past half a century \citep{P81, BH98, AF13}. The complex radiative features observed from these systems such as outbursts of dwarf novae \& low mass X-ray binaries \citep{W03, S00, LV06, BZ15} and aperiodic variabilities of X-ray binaries \citep{RM06, LV06, B10} \& AGNs \citep{PS88, GM99, IC09, S18} indicate the existence of different limits of accretion. Temporal evolution of radiative output of the accretion systems are attributed to different phases of inner sub-Keplerian accretion flow such as Advection Dominated Accretion Flow (ADAF) \citep{NY94, NY95, A95, FBG04, B05, RM06, KB15}, Convection Dominated Accretion Flow (CDAF) \citep{IA99}, adiabatic inflow-outflow solution (ADIOS) \citep{BB99}, luminous hot accretion flow (LHAF) \citep{Y01}, jet \citep{R03, BAP05, P12} corona \citep{GL05, G10, KGS15} etc, fed by the outer standard cool Keplerian flow \citep{SS73, NT73}. Each of these phases of the accretion flow is characterised by an appropriate $\alpha$ parameter and cooling mechanisms \citep{YN14} determined by mass accretion rate. 

From the spectral analysis of AGNs and X-ray transients, now it is fairly understood that the accretion discs around black holes and neutron stars are hot optically thin geometrically thick flow domain surrounded by a much larger optically thick geometrically thin disc. In the transition region these two flow domain can coexist where a cool disc is embedded in a hot coronal flow \citep{WNX15, IK19, GVR21}. Such a structure raises the possibility of interaction of cool disc and hot corona. The radiative transport of the disc will be affected by the presence of hot corona and the global mass-energy exchange of the disc-corona system is modelled separately \citep{MHLQ17, HM91, MLM00, L15, QL17}. Moreover, the hydrodynamic and magnetic components of the viscous stress could seriously be modified in shear flow with such a strong vertical temperature jump. Therefore it is important to analyse the temporal evolution of the local shearing patch corresponding to the disc-corona system and study the exchange of matter and magnetic energy.

The disc-corona interaction is fairly universal and many phenomena associated to accretion systems can be reduced to such a topology. In the rest of the introduction we briefly identify a few such cases. Interaction between the outer disc and the radiation originating from the very inner part of the disc was suggested in the context of outburst in dwarf nova and X-ray transients \citep{PM95, L99}. The thermo-viscous instabilities causing these outburst are modelled as influenced by this radiation exposure \citep{PM94, KR98, L01}. Later on it was shown that \citep{RC96, R99, MKZ97} the heating of the outer disc from top by the radiation originating from the inner part of the disc could create a hot corona above the cool disc. A transition region of sharp temperature gradient between
the hot coronal layer of temperature $\sim 10^9$ K and the disc of temperature $ \sim 10^4$ K is created. Such a configuration is shown to be unstable which will spontaneously create hot coronal clouds above the cool disc. The interaction between the disc and the hot coronal patches could influence the evolution of the disc properties particularly when they are magnetically coupled. 

The physical mechanisms operating in accretion systems of different scales such as X-ray binaries and quasars are similar in nature. However since the sources of the matter inflow are different, the very outer part of the discs themselves may have different topologies. In the case of X-ray binaries the matter is supplied by Roche Lobe overflow and hence the outer disc is mostly concentrated about the midplane of the disc. In the case of quasars or AGN's the matter is sourced by the hot wind from surrounding stars and hence the outer disc itself could have a relatively denser corona \citep{L15, QL17}. Hence in this context also the interaction between the disc and corona could play crucial role in the dynamical evolution of the accretion system.  

Magneto Rotational Instability (MRI) \citep{C60, BH91} and origin of turbulent viscosity in accretion disc has been discussed by numerous authors particularly with the types of initial magnetic field configurations such as an effective poloidal magnetic field \citep{HGB95, BS13, SA16}, a zero net initial magnetic field or a weak toroidal field \citep{BNS95, MS00, DSP10, SBA12}. The measured value of $\alpha$ viscosity parameter in numerical experiments is smaller than at least by an order of magnitude of the expected value \citep{KPL07}. Various modifications and generalisations such as gravitational stratification, radially extended flow domain and effect of radiative transport were explored by different authors \citep{MS00, HKS06, WBH03}. Numerical experiments on the vertically extended disc could generate magnetically dominated corona above a gas pressure dominated disc \citep{SA16, KGS18} for sufficiently strong initial magnetic field. Heating due to magnetic reconnection \citep{LMO03, HWW14} in this magnetically dominated corona could create a sharp temperature gradient across ``the cool disc -- hot corona system''. 

For all the plausible scenarios mentioned above, a vertically extended disc with temperature jump symmetrically above and below the mid-plane is a simple and faithful representation of the actual physical problem. On a global scale, the mass exchange between disc and corona, conduction from hot corona to cool disc and the Comptonization of soft photons from disc by hot corona are the main interaction processes \citep{MLM00, RC00, RM06, BM16}. Whereas on a local scale the interesting scenarios are the  growth and evolution of magnetic field topology; hydrodynamic \& magneto-hydrodynamic instabilities and the effective viscous stress produced; and oscillations at the disc corona interface. 

We have studied a local 3D-MHD numerical simulations using a shearing box approximation \citep{HGB95} considering an initial stratiﬁed density distribution due to a linear gravity profile in the vertical direction and strong toroidal magnetic ﬁeld. The disc corona system is mimicked by imposing a temperature jump symmetrically in the vertical direction. This patch of combined disc corona system is allowed to evolve for several rotation times. The paper is organised as follows:
in \Sec{model} we the setup of our model, in \Sec{res} we
have presented the results. We then discuss our findings and
conclude in \Sec{dc}.
  
\section{Model \& Numerical Setup}
\label{model}

We numerically model a local patch of an accretion disc with
a hot corona above by solving fully compressible hydromagnetic
equations using the publicly available
\textsc{Pencil Code}\footnote{\url{http://github.com/pencil-code}}
which is a high-order, finite-difference, modular, MPI code.
Basic equations being solved may be expressed as:
\begin{align}
\frac{D \ln \rho}{Dt} &= -\bm\nabla\cdot\bm{v}, \\
\frac{D\bm{v}}{Dt} &= \bm{g} +\frac{1}{\rho}
\left(\bm{J}\times\bm{B}-\bm\nabla p
+\bm\nabla \cdot 2\nu\rho\bm{\mathsf{S}}\right)
-2\Omega_0 \hat{z}\times\bm{v} \\
\rho T\frac{D s}{Dt} &= 2\rho \nu \bm{\mathsf{S}}^2 + \mu_0\eta\JJ^2 -\bm\nabla\cdot\bm{F}_{\rm SGS}
- (\gamma-1) \rho c_p \frac{T-T_{\rm d,c}}{\tau_{\rm c}}\, ,
\label{equ:ss} \\
\frac{\partial \bm A}{\partial t} &= {\bm v}\times{\bm B}
- \eta \mu_0 {\bm J},
\end{align}
where $\bm{v}$ is the velocity,
$D/Dt = \partial/\partial t + \bm{v} \cdot \bm\nabla$ is the
advective time derivative,
$\bm{g}$ is the gravitational acceleration with a vertically linear profile, $\hat{z}$ is the unit vector along the
vertical $z$-direction,
$\mathsf{S}_{ij}=\half(v_{i,j}+v_{j,i})
-\onethird \delta_{ij}\bm\nabla\cdot\bm{v}$
is the traceless rate of strain tensor, where commas denote
partial differentiation,
$\nu=\const$ is the kinematic viscosity,
$s$ is the specific entropy, $\rho$ is the fluid density,
$p$ is the pressure,
${\bm A}$ is the magnetic vector potential,
${\bm B} =\bm\nabla\times{\bm A}$ is the magnetic field,
${\bm J} =\mu_0^{-1}\bm\nabla\times{\bm B}$ is the current density,
$\eta=\const$ is the magnetic diffusivity,
$\mu_0$ is the vacuum permeability which is taken to be unity
in our units,
$\gamma=\cp/\cv$ is the ratio of specific heats
at constant pressure and density, respectively,
and $T$ is the temperature.

In some cases we have also adopted a Smagorinsky model for
viscosity with $\nu=\nu_S$ where
\begin{equation}
\nu_S=(C_k\Delta)^2\sqrt{\bm{\mathsf{S}^2}}\,.
\end{equation}
Here, $C_k$ is the Smagorinsky constant and $\Delta$
is the filtering scale which is chosen to be equal to the grid spacing; see, e.g., \cite{HB06}.
We have used $C_k=0.35$ in this work.

The last term in \Eq{equ:ss} is a relaxation term which
guarantees that the temperatures in the two subdomains,
disc and corona, remain, on average, constant and
equal to $T_{\rm d}$ (disc) and $T_{\rm c}$ (corona),
respectively.
In the present work, we applied the relaxation term only in
the corona to maintain its temperature. This is preferable
as it allows the flow to evolve more freely in the disk.
In one of the simulations, A1s$\chi$, we have used a subgrid-scale (SGS)
diffusivity ($\chi_{\rm SGS}$) which acts on the
fluctuations of the entropy about its
horizontal average \citep{Kap21}. The SGS flux is given by
\beq
\bm{F}_{\rm SGS} = -\rho T \chi_{\rm SGS} \bm\nabla s'\;,\quad
s'=s-\langle s \rangle_{xy}
\eeq

The angular velocity of the accretion disc at some
arbitrary radius $R=R_0$ is denoted by $\Omega_0$.
In rotating reference frame of the local Cartesian
patch of the disc at $R_0$, the velocity field is in the
toroidal $y$-direction with a linear shear profile:
\beq
\bm{V} = -q \Omega_0 x \,\hat{y}
\eeq
where $q=3/2$ for a Keplerian disc, and $\hat{y}$ is
the unit vector along toroidal direction. The total velocity field is $\bm{v} = \bm{V} + \bm{u}$, 
where $\bm{u}$ is the velocity deviation.

\begin{figure*}
\centering
\includegraphics[width=\columnwidth]{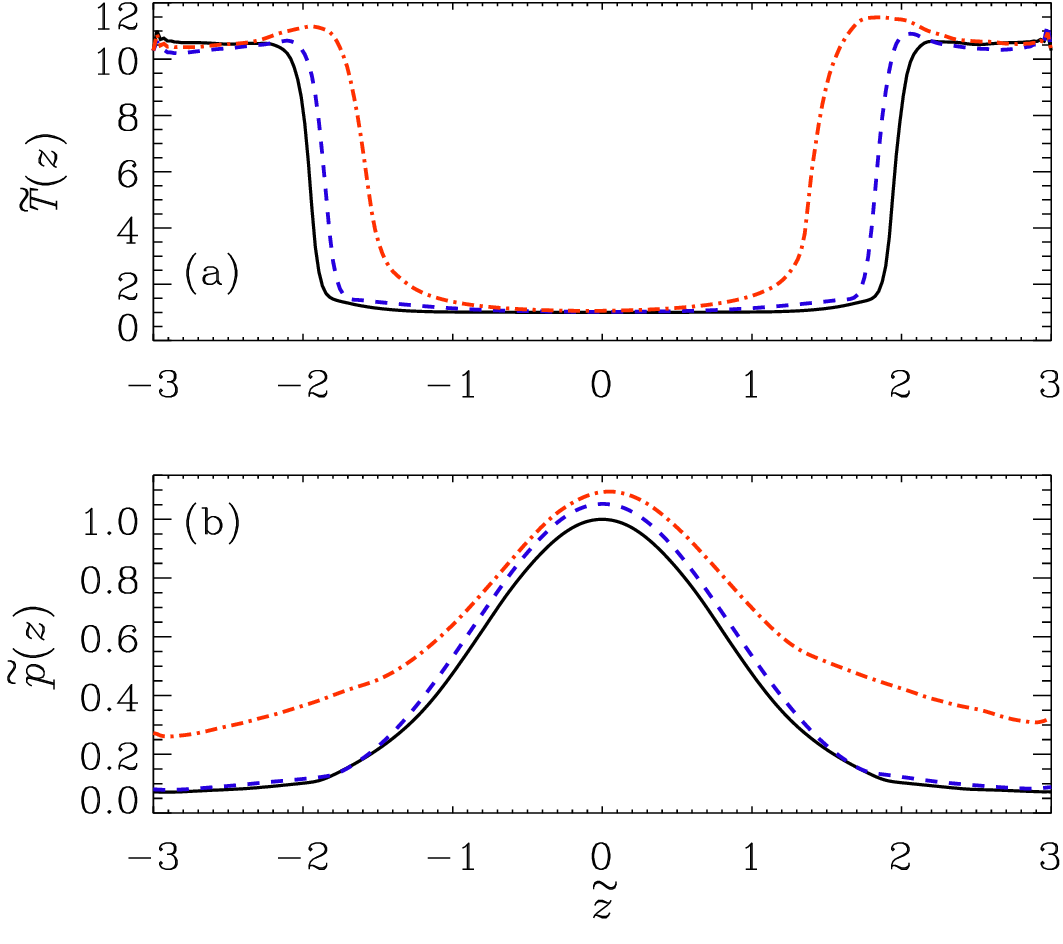}
\includegraphics[width=\columnwidth]{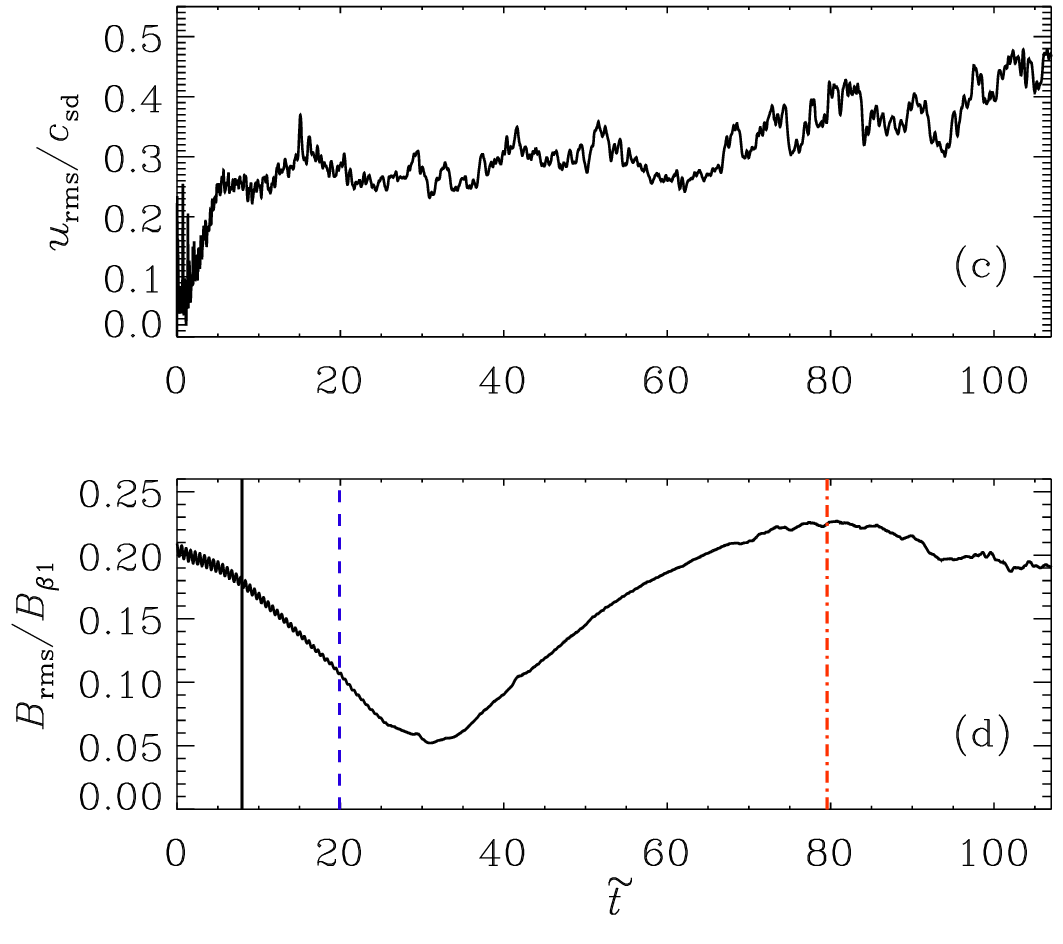}
\caption{Profiles of (a) temperature, and (b) pressure, as functions of
normalized vertical coordinate $\tilde{z}$, with $\tilde{z}=0$ being the
midplane of the disc. Panels (c) and (d) display the temporal evolution
of root-mean-squared values of total fluid velocity and total magnetic
field, respectively. Solid (black), dashed (blue) and dash-dotted (red)
curves in (a) and (b) show the profiles at three epochs in time that
are marked by corresponding vertical lines in panel (d).
These are shown from the run A1s$\chi$ as listed in \Tab{tbl1}.
}
\label{profs1} \end{figure*}

We assume an
ideal fluid with an equation of state determining its
pressure by $p=(\cp-\cv) \rho T = \rho \cs^2/\gamma$,
where $\cs$ is the adiabatic sound speed. Note that we have
a piecewise isothermal setup where disc (with sound speed $\csd$)
and a hotter corona (with sound speed $\csu$)
are maintained at two different temperatures, as shown,
for example in \Fig{profs1}a. The sharp jump in temperature or
density at the disc-corona interface in the beginning is characterized by
\beq
\delta_0 = T_{\rm c}/T_{\rm d} = \csu^2/\csd^2 = 
\rho_{i-}/\rho_{i+}\,,
\eeq
where $\rho_{i-}$ and $\rho_{i+}$ are the densities just
below and above the interface, respectively. 

\begin{table*}[t!]\caption{Summary of runs. All runs have
$H=1$, $\sh=-1.5$, $q\equiv-S/\Omega_0=1.5$, and $\delta_0=10$.
}
\centerline{
\begin{tabular}{lcccccccc}
\hline\hline\\[-2mm]
Run & Domain & Grid & $\beta_{0}$ & $\ma$ & $\re$ & $\rem $ & Viscosity \\
\hline
A1s$\chi$ & $6H\times6H\times6H$ & $256\times256\times256$ & 5.7 & 0.28 & -- & 70.5 & Smagorinsky\\
A2s & $6H\times6H\times6H$ & $256\times256\times256$ & 5.7 & 0.31 & -- & 78.2 & Smagorinsky\\
A3e & $6H\times6H\times6H$ & $256\times256\times256$ & 5.7 & 0.16 & 78 & 39 & Explicit\\
\hline
B1e & $6H\times6H\times6H$ & $128\times128\times128$ & 1.5 & 0.40 & 134 & 134 & Explicit\\
\hline
C1s & $2H\times2H\times8H$ & $128\times128\times320$ & 3.4 & 0.11 & -- & 1111 & Smagorinsky\\
\hline
\label{tbl1}
\end{tabular}
}
\end{table*}

\subsection{Scaling and initial conditions}

In most simulations we have a local cubical shearing box of
side $6H$ (unless otherwise stated) and angular velocity $\Omega_0$ with a
temperature jump at $\pm2H$ on either sides of the midplane at
$z=0$, where $H=\Omega_0 / \csd$.
Initial density (or pressure) profiles of the medium which is vertically stratified under linear gravity are piece-wise Gaussians in disc and corona. We choose
a Gaussian profile along $\tilde{z}$ for the toroidal field
$B_{y}(z)$ with an initial plasma parameter 
at the midplane being $\beta_0$. With $\urms$ being the
rms value of the fluid velocity $\bfu$, we define dimensionless
quantities as: the plasma parameter as the
initial ratio between the thermal and magnetic pressures,
$\beta_0=p/(B_y^2/2\mu_0)$, where $\beta_0$ is a constant in the
disc region; Mach number $\ma=\urms/\csd$;
fluid Reynolds number $\re=\urms H/\nu$; magnetic Reynolds number
$\rem=\urms H/\eta$; shear parameter $\sh=-q\Omega_0 H/\csd$.

We choose $\Omega_0 = 1$ and $\csd = 1$, giving $H = 1$. Time
and length are scaled by rotation time $T=2\pi/\Omega_0=2\pi$,
and $H$, respectively.
Density is scaled in units of initial midplane density $\rho_0$ and
the magnetic field is scaled in units of $B_{\beta 1} = \sqrt{2 \mu_0 \rho_0}\, \csd$. The scaled magnetic field, time and position are
$\Tilde{B}$, $\Tilde{t}$ and $\left(\Tilde{x}, \Tilde{y}, \Tilde{z} \right)$, respectively.  

\subsection{Boundary conditions}
We have used shearing-periodic boundary conditions in the radial direction (ie; $x$ coordinate) that reproduce the differential rotation through the angular displacement of the radial boundaries. The standard periodic boundary condition is applied in the angular direction (ie; $y$ coordinate). We study the interaction and evolution of the two preexisting flow domains namely the cold disc and the hot corona. Hence there is an exchange of matter between them but the combined system conserves matter. Therefore the appropriate boundary condition in the vertical direction (ie; $z$ direction) is the zero outflow condition such that $u_z = 0$ at $z=\pm 3H$. A vertical magnetic field boundary condition is applied at the two boundaries. Besides, the density is extrapolated assuming a vertical hydrostatic equilibrium.

\subsection{Diagnostics}
In order to study the evolution of the system and the associated instabilities we define the following averages. For a quantity
$f=f(x,y,z,t)$, the volume average $\langle f\rangle_V$, and the
planar average $\langle f\rangle$ are given by the expressions,
\begin{equation}
\langle f\rangle_V = \frac{\int f dx dy dz}{\int dx dy dz} \,,
\quad \langle f\rangle  = \frac{\int f dx dy}{\int dx dy}\,. 
\end{equation}
The total stress tensor is given by
\beq
T_{xy}(z,t) = T_{xy}^{\rm Rey}(z,t) + T_{xy}^{\rm Max}(z,t)
\eeq
where $T_{xy}^{\rm Rey}(z,t) = \langle\rho u_x u_y \rangle$
is the Reynold's stress, and
$T_{xy}^{\rm Max}(z,t) = -\langle B_x B_y \rangle / \mu_0$ is the
Maxwell's stress.
The $z$-dependent $\alpha$ viscosity parameter is defined as  
\begin{equation}
\bar{\alpha}(z,t) = \frac{T_{xy}(z,t)}{\langle p\rangle}\,,
\label{alpha}
\end{equation}
where $\langle p\rangle$ is the horizontally averaged gas pressure.
By taking the root-mean-squared (rms) value of $\overline{\alpha}$,
let us define the time-varying viscosity parameter as
$\alpha(t) = \sqrt{\langle \overline{\alpha}(z)^2\rangle_z}$.
The rms values of the fluid velocity and the total magnetic
field are given by $\urms=\sqrt{\langle \bfu^2 \rangle_V}$ and
$\Brms=\sqrt{\langle \bfB^2 \rangle_V}$, respectively.
Mean magnetic fields are defined with respect to the
planar averages of normalized magnetic fields as
$\overline{B}_x(z,t) = \langle\widetilde{B}_x \rangle$
and $\overline{B}_y(z,t) = \langle\widetilde{B}_y \rangle$.

\begin{figure*}
\centering
\includegraphics[width=\columnwidth]{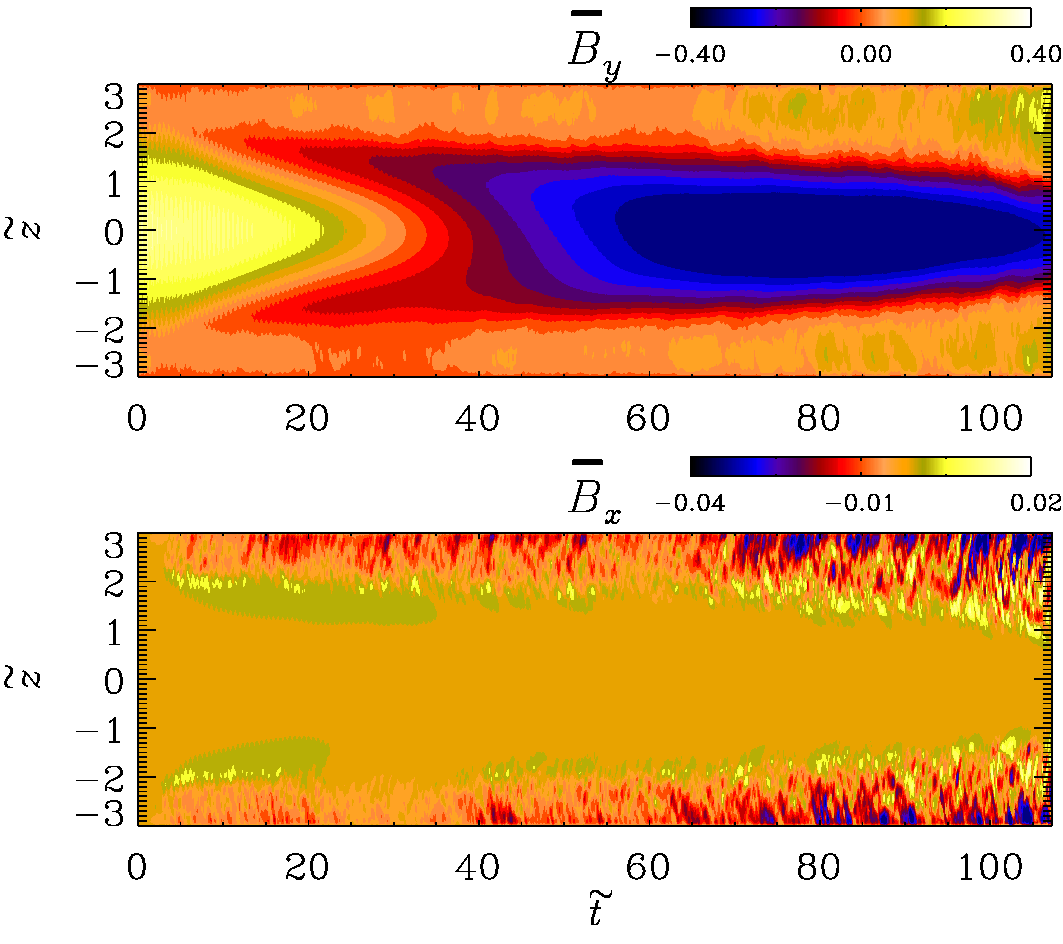}
\includegraphics[width=\columnwidth]{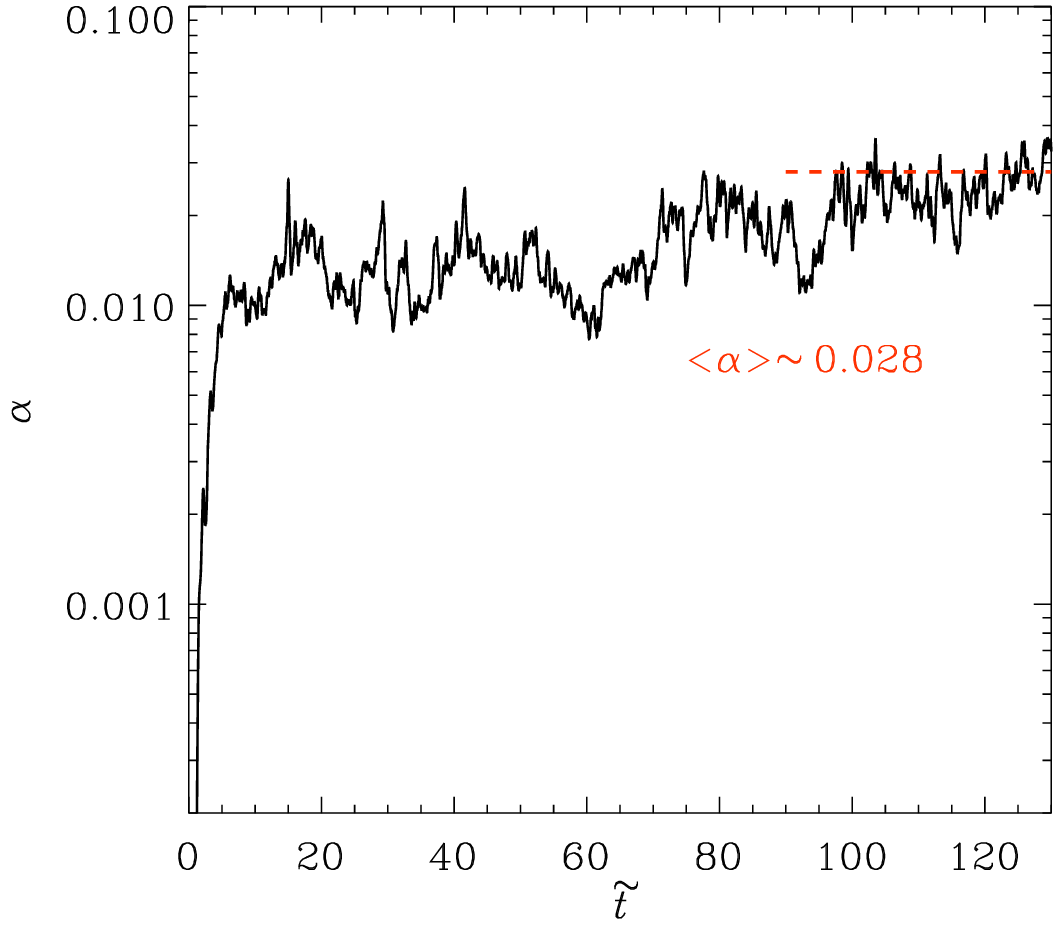}
\caption{Left: spacetime diagrams of horizontally averaged mean magnetic
fields, $\overline{B}_y$ (top) and $\overline{B}_x$ (bottom), which
are normalized by $B_{\beta1}$.
Right: temporal behavior of the viscosity parameter $\alpha$.
These are shown from the run A1s$\chi$ as listed in \Tab{tbl1}.
}
\label{st_alp1} \end{figure*}

\section{Results}
\label{res}
Results of our simulations that are summarised in \Tab{tbl1} are
being presented here.

\subsection{A1s$\chi$ Model}
\label{A1}

This is a large eddy simulation (LES) with a Smagorinsky viscosity in
a shearing box. It covers about $\sim 110$ orbits.
With $\beta_0=5.7$,
initially imposed toroidal magnetic field is strong in this case, and
therefore, the Parker-Rayleigh-Taylor-Instability (PRTI) is expected
to be operational \citep{KGS18}. The run A2s as listed in \Tab{tbl1}
has a slightly larger $\rem$, but is otherwise similar to A1s$\chi$ model.
Findings from these two runs are quite similar, so, below we present
results from the run A1s$\chi$.

\begin{figure*}  
\centering
\includegraphics[width=2\columnwidth]{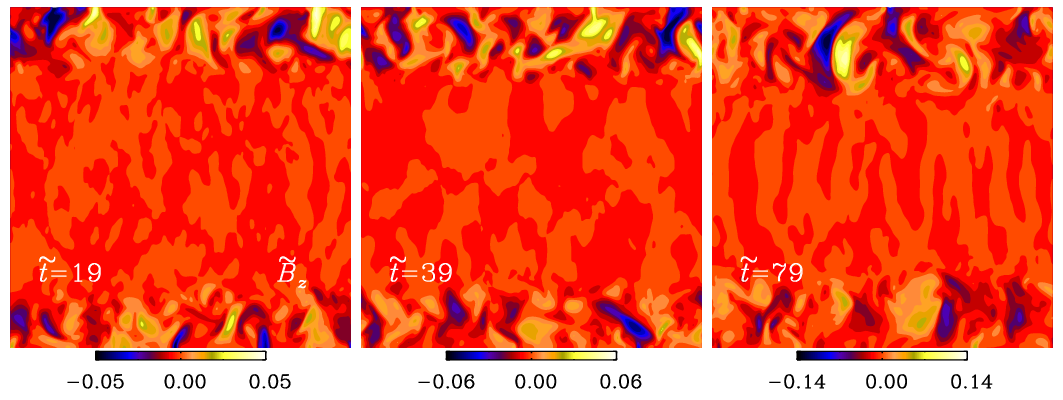}
\caption{Three snapshots of the normalized vertical
magnetic field ($\widetilde{B}_z$) from the $y=0$ plane
from the run A1s$\chi$; see \Tab{tbl1} and \Fig{st_alp1}.
}
\label{Bz} \end{figure*}

\subsubsection{Vertical structure of disc-corona system}

\Fig{profs1} shows the vertical profile of the thermodynamic variables
and the time evolution of volume average based root-mean-squared
(rms) strengths of the fluid velocity ($\urms$) and total magnetic
field ($\Brms$) from
the combined disc-corona system.
As may be seen from \Fig{profs1}a, a temperature jump, with a
corresponding drop in the density (not shown), in the corona by
factor 10 compared to the disc is maintained during the simulation.
Relaxation term in the entropy equation is applied only for the
layers with $|\tilde{z}|\geq2$. As time passes, we do observe a
gradual drift of the transition layer in such a way that the extent of
the corona increases. This is expected due to heating in the system
by magnetic reconnection.

Thus we a have a piece-wise isothermal domains where a cold disc
is embedded between coronal envelope of higher temperatue.
System is vertically stratified under liner gravity which
leads to correspondingly piece-wise Gaussian profiles for pressure
as well as density in the disc and corona; see \Fig{profs1}b where
$\tilde{z}=0$ is the midplane of the disc, and the pressure is continuous across the interface.

Turbulence is quickly produced within a few rotation time due to the
magnetic field. Flow is subsonic and $\urms$ shows a saturation after
about 20 rotation time. The simulation starts with an initial toroidal
magnetic field $B_y$ which is Gaussian in structure with $\beta_0=5.7$.
\Fig{profs1}d shows that $\Brms$ decays in the beginning and after about
30 rotation times, it starts to grow. This growing phase is linked to
the reversal of the toroidal magnetic field; see the paragraph below
for a discussion on the reversal of $\Bym$. Both, the magnetic field and
the turbulence, are maintained self-consistently in this system.

\subsubsection{Reversal of toroidal magnetic field ($\Bym$)}

\Fig{st_alp1} shows the spacetime diagrams of the
horizontally averaged mean magnetic fields, $\Bym$ (left, top)
and $\Bxm$ (left, bottom), which are normalized by $B_{\beta1}$ defined
earlier. Remarkably, the mean toroidal magnetic field undergoes a
complete reversal in time by changing its sign, and it is predominantly
confined within the disc. This is a rather unique class of evolution
of the magnetic field which has not been reported earlier;
see the spacetime diagrams in \Fig{st_alp1}, and in a number of other
cases discussed below. The $x$-component of the mean magnetic field is
mostly confined in the coronal regions, and it is
generated by the MHD instabilities in this system.

Vertically migrating dynamo waves are commonly seen in a number
of previous works which typically model an isothermal disc
\citep{BNS95, SBA12, SA16a, KGS18}. The reversal of the toroidal magnetic field that we find in this work is quite intriguing.
Unlike magnetic field solutions in an isothermal boxes considered
in earlier works, we find here that the first moments of the magnetic
fields, $\Bxm$ and $\Bym$, are spatially separated. This is caused by
the hot corona above the disc.

\subsubsection{Vertical magnetic fields}

In \Fig{Bz} we show profiles of the unaveraged vertical magnetic fields
from the snapshots taken at different times where we chose to display
the structure from the $y=0$ plane. These fields are produced by the
action of buoyancy effects in such stratified systems which
are expected to host instabilities such as MRI and PRTI.
As is evident from \Fig{Bz}, $B_z$ is predominantly confined in the
low density coronal regions where it appears to be of small scale
in nature and its strength increases in time.

\subsubsection{Viscosity parameter ($\alpha$)}

Time evolution of the Shakura-Sunyaev (SS)
viscosity parameter ($\alpha$) as defined
below \Eq{alpha} is shown in the right panel of \Fig{st_alp1}.
We find that $\alpha$ saturates with a mean value of $\sim 0.03$
after about 50 rotation times. This is about ten times larger compared
to the value of $\alpha$ obtained in \citet{BNS95}, and close to the
one seen in simulations of \citet{KGS15}.

\begin{figure*}  
\centering
\includegraphics[width=\columnwidth]{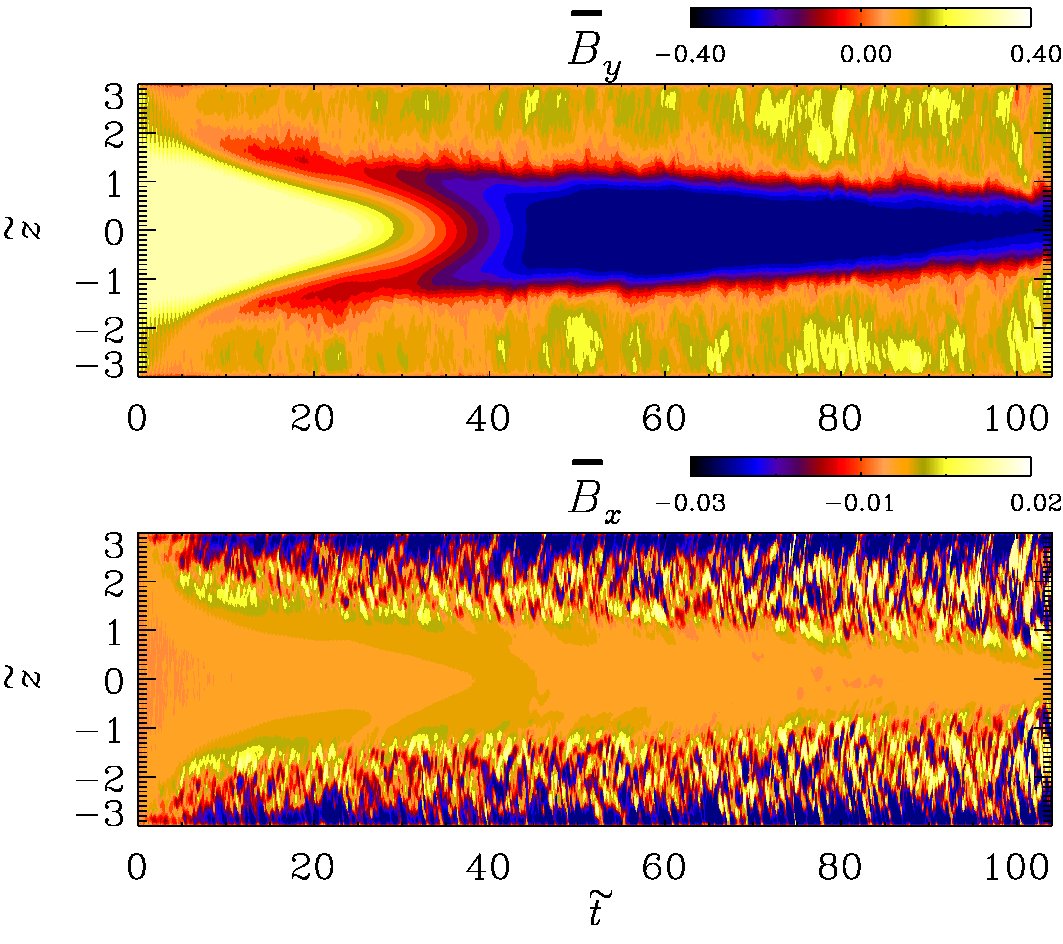}
\includegraphics[width=\columnwidth]{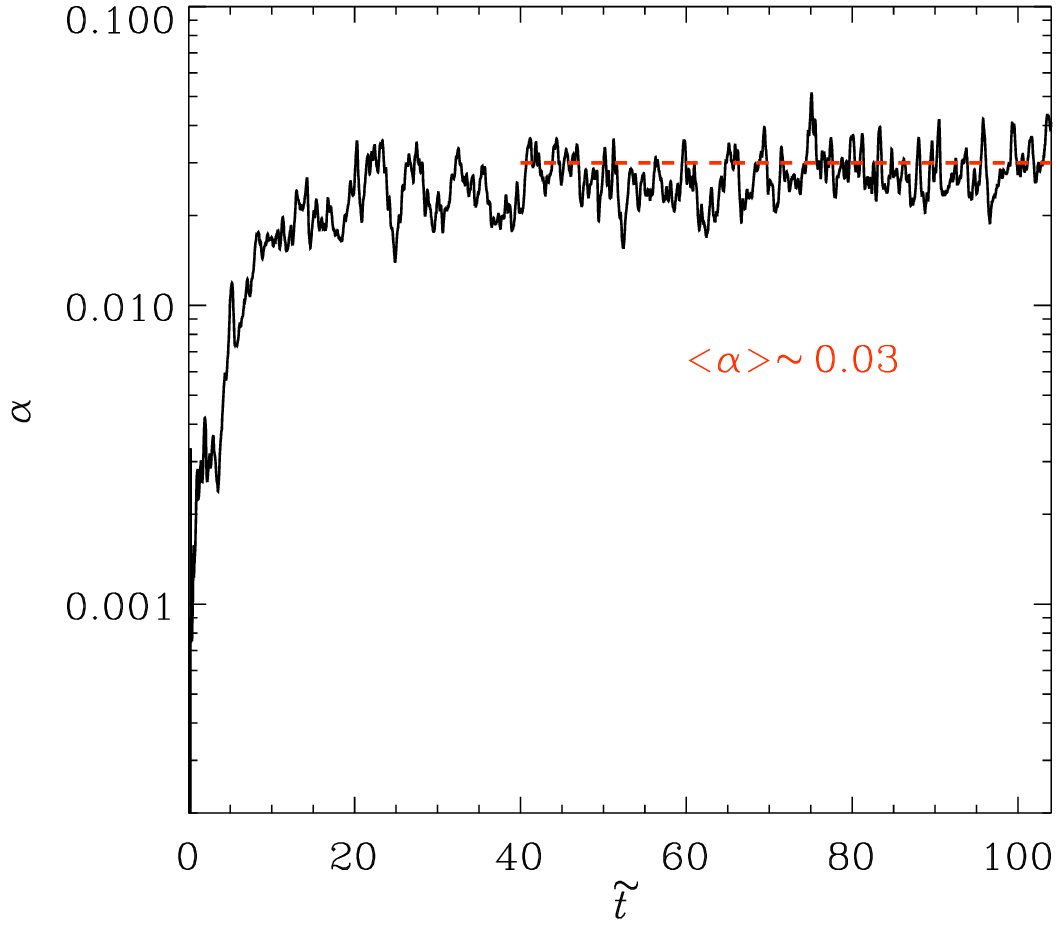}
\caption{ 
Same as \Fig{st_alp1} but from the run B1e listed in \Tab{tbl1}.
}
\label{st_alp2} \end{figure*}

\begin{figure*}  
\centering
\includegraphics[width=\columnwidth]{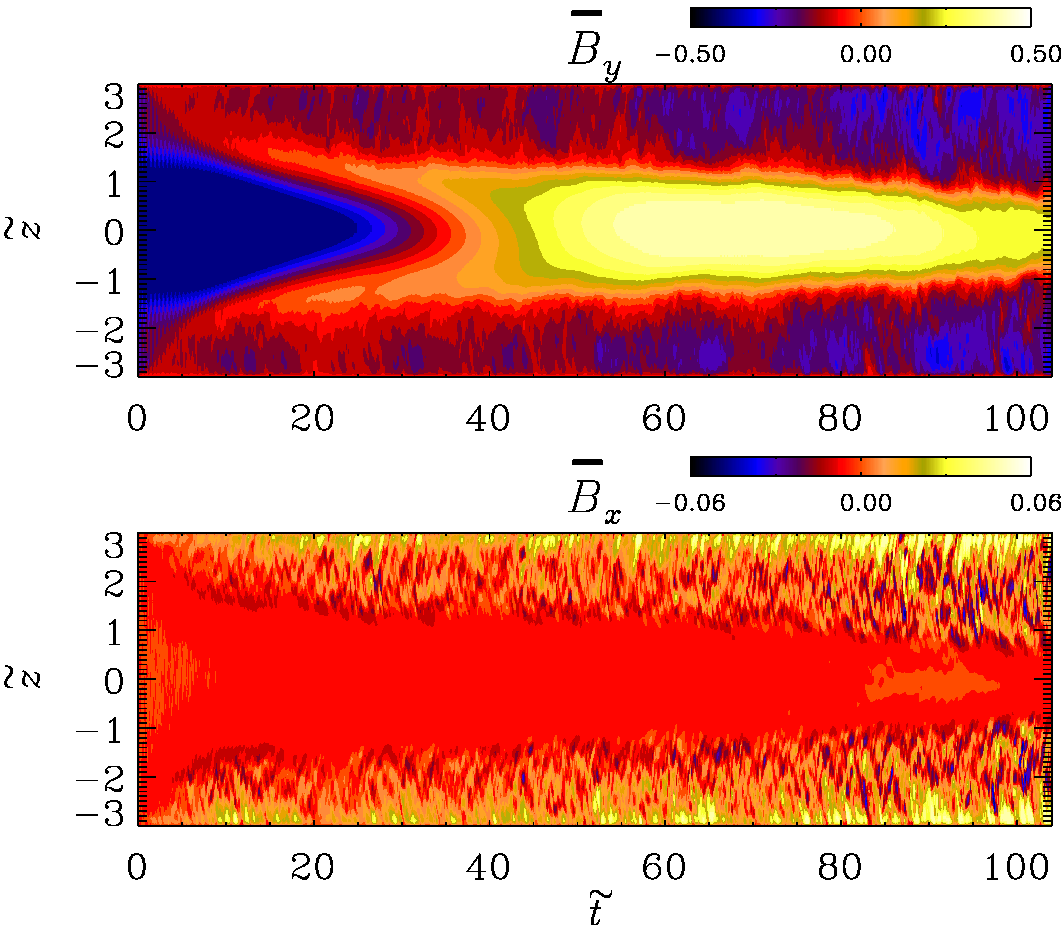}
\includegraphics[width=\columnwidth]{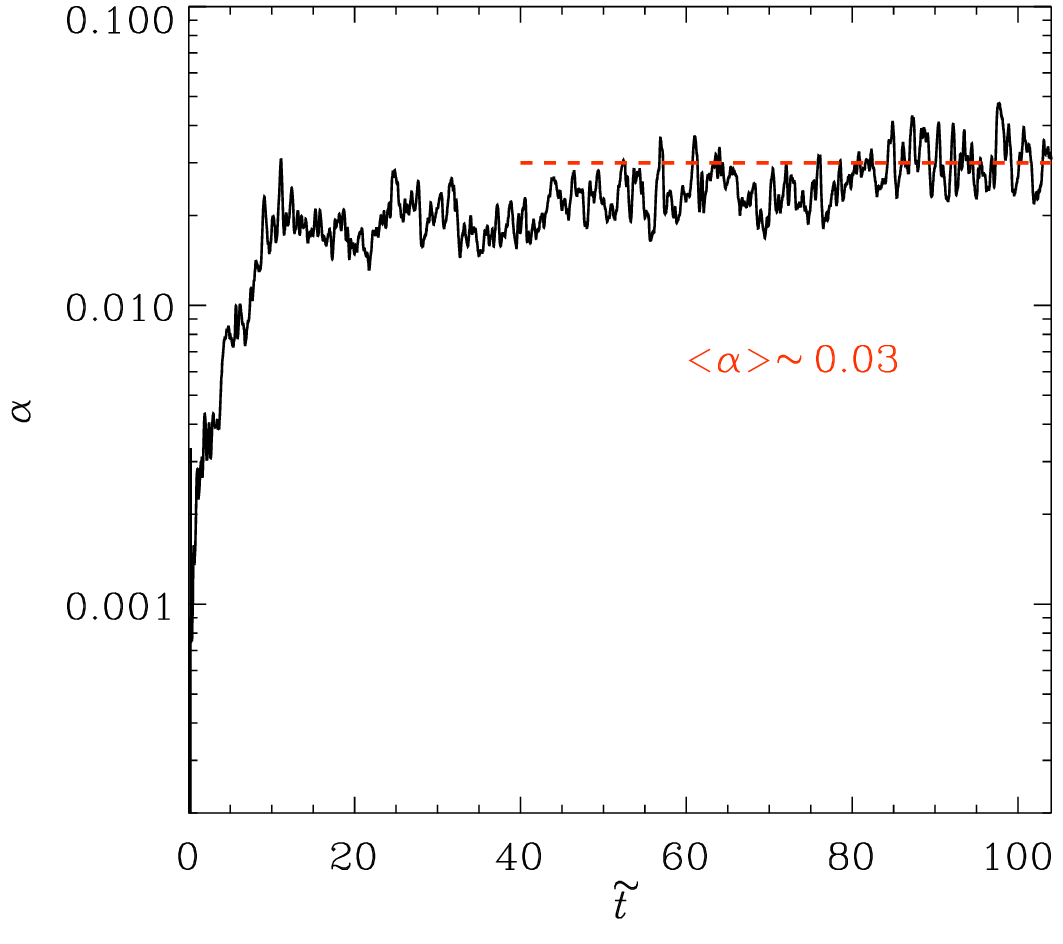}
\caption{ 
Same as \Fig{st_alp1} but from the run ``B1eN''
(not listed in \Tab{tbl1}) which is identical to the run B1e of
\Tab{tbl1}, except that the initial $B_y$ in this case has the opposite
sign.
}
\label{st_alp3} \end{figure*}

\begin{figure*}  
\centering
\includegraphics[width=\columnwidth]{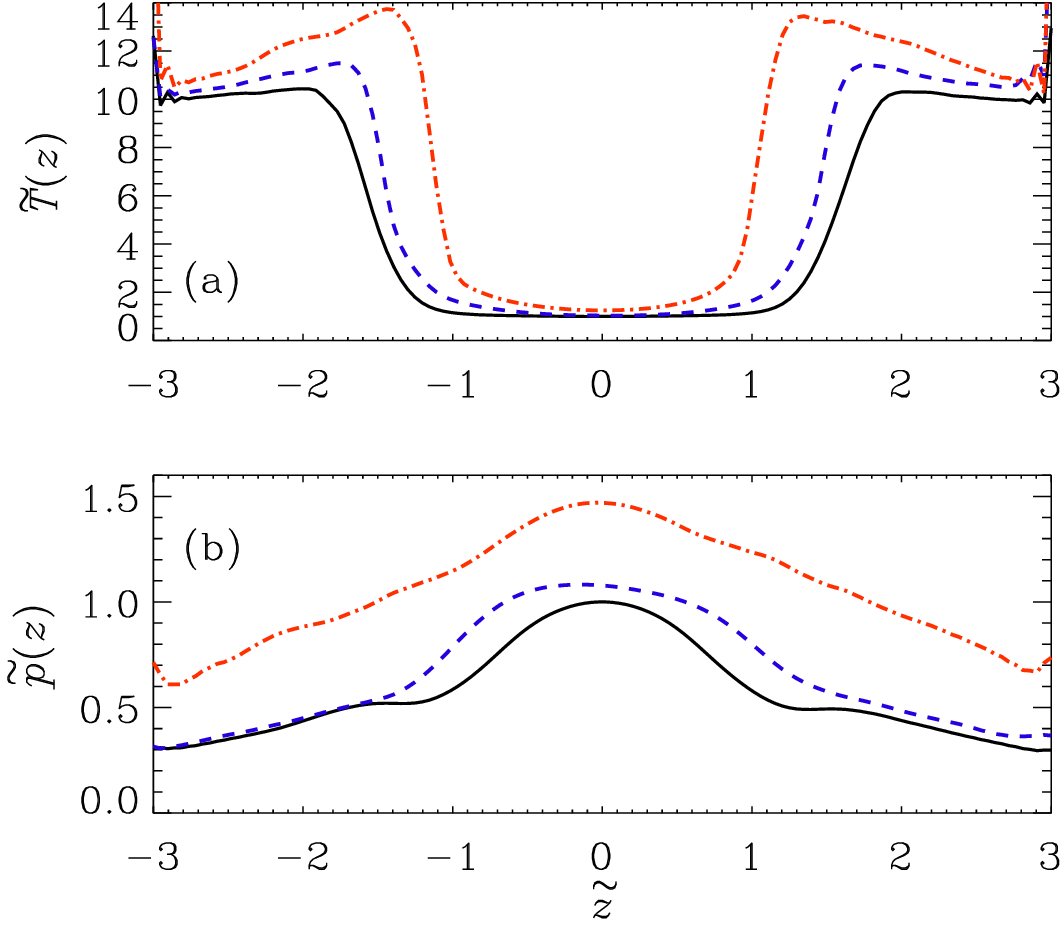}
\includegraphics[width=\columnwidth]{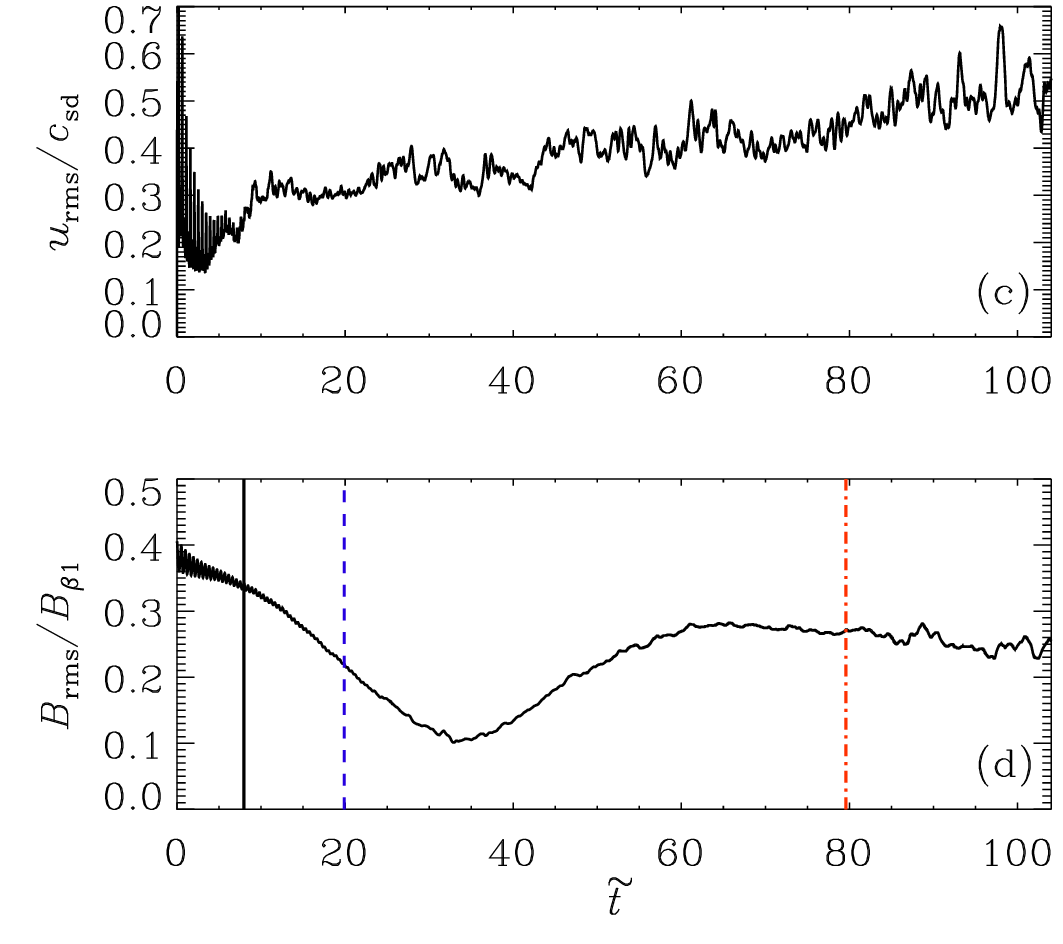}
\caption{ 
Same as \Fig{profs1} but from the lower resolution run ``B1eN''; see
the caption of \Fig{st_alp3}.
}
\label{profs2} \end{figure*}

\subsection{Comparison between models with different numerical
schemes for kinematic viscosity ($\nu$)}

In order to test how sensitive are our results to the numerical schemes
adopted for $\nu$, we perform another
set of simulations
with explicit kinematic viscosity. One such run, called A3e in \Tab{tbl1},
is presented in this work; see \App{expnu} where we include results from
this run with explicit $\nu$.
Note that this has smaller $\rem$ (and $\re=78$) compared to
the A1s$\chi$ model with Smagorinsky viscosity discussed above in
\Sec{A1}. Broad conclusions from models with these two different
numerical schemes for $\nu$ are same, i.e., the toroidal magnetic
field undergoes a complete reversal as discussed in \Sec{A1}. Further
discussion on this deferred to \App{expnu}.

\subsection{Models with $\beta_0=1.5$ and oppositely directed initial
magnetic field configurations}

Here we present results from two lower resolution runs, B1e (listed
in \Tab{tbl1}), and B1eN which is identical to the run B1e, except that
the initial $B_y$ in this case has the opposite sign. We have
used explicit $\nu$ in both these cases where we test the robustness
of our findings discussed above on the field reversals.
In \Figs{st_alp2}{st_alp3} we show the spacetime diagrams of mean
magnetic fields, $\overline{B}_x$ and $\overline{B}_y$, as well as
the time evolution of the SS viscosity parameter $\alpha$.
Mean toroidal field $\overline{B}_y$ undergoes a complete reversal
in time by changing its sign in both these cases, and the SS viscosity
parameter $\alpha$ saturates to a mean value of $\sim 0.03$ after about
30 rotation times.

As may be seen from \Fig{profs2}(a), there is a gradual drift of the
disc-corona interface also in these cases such that the
extent of hotter corona slowly increases. Recall that the relaxation
term, as discussed in \Sec{model}, to maintain the corona of higher
(lower) temperature (density) is applied only in layers with
$|\tilde{z}|\geq2$. \Fig{profs2}(c) shows that the instabilities in the magnetized disc drive turbulence, which, in turn, governs the
evolution of magnetic field in a self-sustaining manner.
Ohmic heating due to magnetic reconnection of small scale loop-like
structures which are prominant in the low density coronal regions
is likely the cause for further heating of corona. This is expected
to play a crucial role in the formation of corona in the magnetized
accretion discs.

\subsection{A discussion on the run C1s in a tall box}

We have also performed some simulations in a tall box where the
vertical extent is four times larger than the horizontal extents;
$\tilde{z}\in [-4,4]$ with relaxation term to maintain the corona
operating in $|\tilde{z}|\geq2$ layers. Corona is therefore
thicker compared to the cubic domains considered in cases above.
Due to piece-wise isothermal setups that are vertically stratified
under linear gravity being studied in this work, the densities become
vanishingly small in top layers of corona in such tall boxes.
This leads to numerical challenges and increases the computational cost
of such models. Nevertheless, some early findings from this run seem
interesting enough, so we mention them briefly here without
showing any plots from this run.

As in other cases discussed above, the run begins with a toroidal
field $B_y$ with a Gaussian profile along $\tilde{z}$ such that
the initial $\beta_0=3.4$ is constant in $\tilde{z}$.
Buoyancy instabilities such as PRTI lead to the generation of
other components of magnetic fields in a manner similar to the
one discussed in \citet{KGS18}. Corona acts as a reservoir of
small-scale loop-like magnetic fields, especially the $B_z$.
Ohmic heating due to magnetic reconnection as studied by \citet{KGS18}
is expected to be more efficient in this run with a thicker corona.
As we enforce a constant temperature at the vertical boundaries
determined through the relaxation term in \Eq{equ:ss}, the excess heat
produced is therefore pushed towards the cool disc. This makes the
combined system asymptotically isothermal over $\sim100$ rotation times.
The resulting solutions for the mean magnetic fields, $\overline{B}_x$
and $\overline{B}_y$, thus display the well known vertically migrating
butterfly patters found in a number of previous works
\citep{BNS95, SBA12, SA16a, KGS18}.

\section{Discussion \& Conclusions}
\label{dc}

The geometrically thin gas pressure dominated accretion disc cannot completely explain the phenomenology of accretion disc around compact objects such as X-ray Binaries and AGNs \citep{Mis20}. Thermo-viscous instabilities and high energy spectral characteristics from these sources indicate that hot geometrically thick and magnetised component of accretion flow is also present \citep{GJL12, BP07}. Magnetic energy dissipation is the main internal mechanism of heating of the disc except the irradiation by very inner part of the accretion disc. There are several attempts to explain the self-sustained generation of large scale magnetic field in accretion around compact objects. It is argued that even if the system starts with a very week magnetic field, strong toroidal magnetic field could be sustained in a later phase \citep{KU02, PP05} driven by MRI \citep{MS00, KGS15, BAP15}. In our work we choose such an initial organised toroidal magnetic field \citep{BP07}, which is superposed on a preexisting
cold disc -- hot corona, as discussed in the introduction. 

Main results of this work can be summarised as follows:
\begin{enumerate}
\item [(a)]The cold disc-hot corona vertical temperature profile is stable: in all the simulations the symmetric step profile of the temperature as a function of vertical coordinate is maintained. Results are independent of Smagorinsky or explicit scheme for the kinematic viscosity. 
\item [(b)] Instabilities in the magnetized disc drive turbulence, which, in turn, governs the
evolution of magnetic field in a self-sustaining manner.
\item [(c)] The large scale toroidal magnetic field is largely confined to the cold disc region: contrary to the general belief, in cold disc-hot corona system, the large scale toroidal magnetic field is strong in the disc region and weakens in the corona region. It is found that the system suppresses the magnetic buyount force, confining the large scale toroidal magnetic field in the disc region. 
\item [(d)] Remarkably, the mean toroidal magnetic field undergoes a
complete reversal in time by changing its sign, and it is predominantly
confined within the disc. This is a rather unique class of evolution
of the magnetic field which has not been reported earlier.
Toroidal magnetic field thus shows an aperiodic field reversal.

Vertically extended shearing box simulations in other
works where an isothermal gas is modeled, one typically
finds dynamo waves, i.e., a butterfly pattern for mean
fields which show quasi-periodicity over time scales
on the order of 10 rotation times. In our case with
a piece-wise isothermal disc-corona system, we find that
the toroidal fields reverses over a time scale of the order of $50$ rotation times, and the commonly seen butterfly pattern of the dynamo
wave is absent.
\item [(e)] Saturated value of the SS viscosity parameter
$\alpha$ is about $0.03$ for several tens of rotation
times.
This is in conformity with most of the shearing box simulations of magnatised accretion disc. The Maxwell's stress is found to be stronger than Reynold's stress and it largely contributes
to $\alpha$.
\end{enumerate}

Thermal Comptonization \citep{SLE76, ST80} in the hot and low dense corona is generally invoked to explain the high energy emission. Apart from the continuous corona, distinct active regions of coronal clouds above the disc are also suggested in this context \citep{HMG94}. In a magnetically coupled disc-corona system, the interaction could cause further increase of the coronal temperature \citep{MF01, K90, Z90,HM91}. The amplification of seed magnetic field in disc by differential rotation and convection is generally balanced by microscopic diffusivities. In a gravitationaly stratified gas, a horizontal magnetic field can trigger unstable modes leading to Parker instability \citep{pr58, pr66}. In an isothermal setup if the magnetic field cannot dissipate at the rate of amplification then magnetic field could emerge from the disc to coronal region due to buoyoncy and magneric loops could dissipate in the coronal region \citep{GRV79}.

Differentially rotating systems such as disc galaxies show
magnetic field reversals while moving radially
\citep{V96, F01, VE11}. Mechanisms such as turbulent dynamo effect and stellar feedback are involved to understand the evolution of magnetic fields in these systems \citep{BLG15,KLD09}. Fully isothermal, shearing box
simulations to model accretion discs reveal a butterfly
pattern where the dynamo wave leads to a quasi-periodicity
in mean magnetic fields with a period of about 10 rotation
time \citep{BNS95, KGS15, SA16a, KGS18}.
Whereas in our two-layer system of cool disc and a hot corona, the quasi-periodicity of mean fields as found in earlier works
is absent. Instead, the mean toroidal magnetic field undergoes a
complete reversal over a time scale of about
50 rotation by changing its sign, and it is predominantly confined within the disc.

Interaction of a hot corona on top of a cool disc was invoked in the context of the so called slab model to explain 
thermal Comptonized X-rays from the disc \citep{DJ97, DJ2000}. Observations in Black Hole Binaries and AGN's show that the fraction of total energy dissipated in corona is very large \citep{HM91, PET18}. Magnetically supported steady state accretion disc corona models with active MRI \citep{GR19} also suggest that an appreciable amount of energy release could happen in the corona region.

As noted above, $\overline{B}_y$ is largely confined to the
disc region due to the presence of a hot corona above. Whereas
$\overline{B}_x$, and also $B_z$ are prominent in the corona;
see \Sec{A1}. We envisage that Ohmic heating due magnetic
reconnection of smaller scale field structures will be more
efficient in the coronal region, producing thus an excess heat
which may heat-up the disc. Corona thus `advances' towards the disc, swallowing the matter from the disc to facilitate more efficient accretion of the matter. This particular situation could creates a scenario of accretion via corona.

\section*{Acknowledgments}
This work used the High Performance Computing Facility of IUCAA, Pune (\url{http://hpc.iucaa.in}). 
AA thanks the Council for Scientific and Industrial Research, Government of India, for the research fellowship. 
SRR thanks IUCAA, Pune for the Visiting Associateship Programme.

\appendix
\section{Field reversals in runs with explicit kinematic
viscosity}
\label{expnu}

\begin{figure}[h!]  
\centering
\includegraphics[width=0.48\columnwidth]{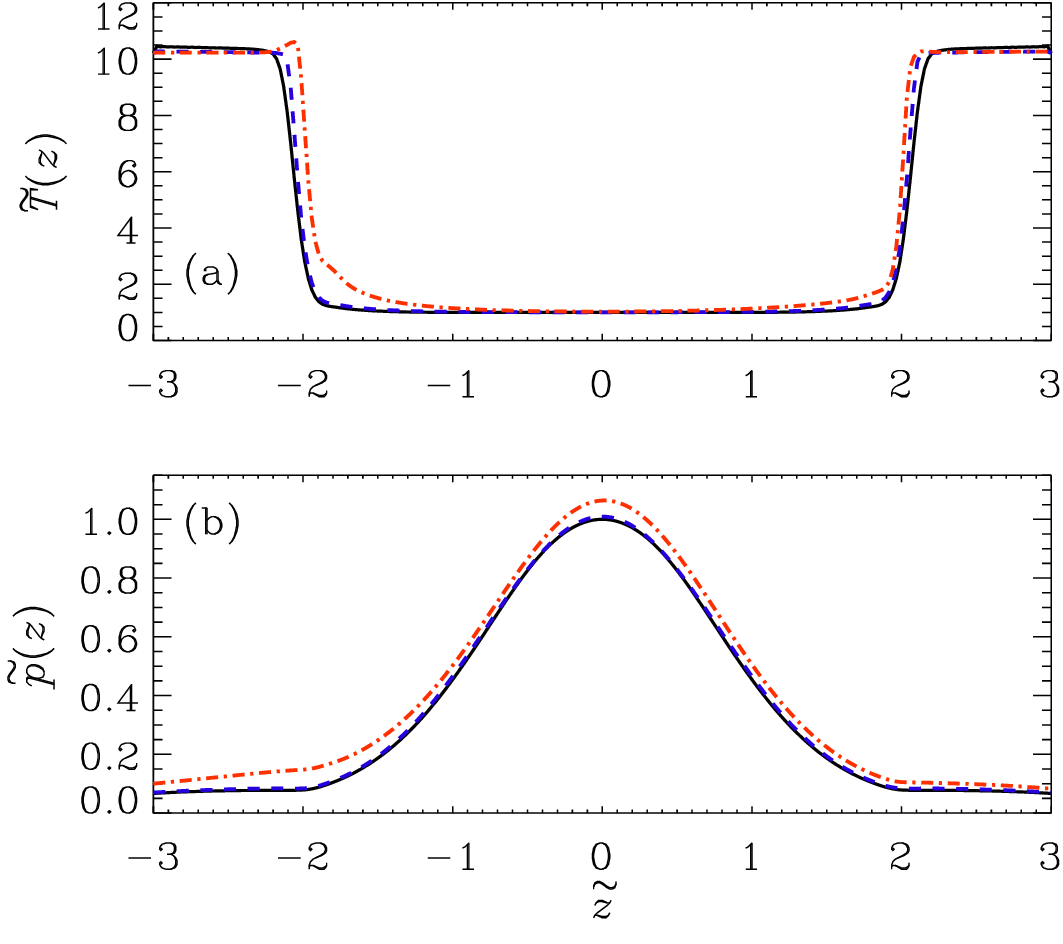}
\includegraphics[width=0.48\columnwidth]{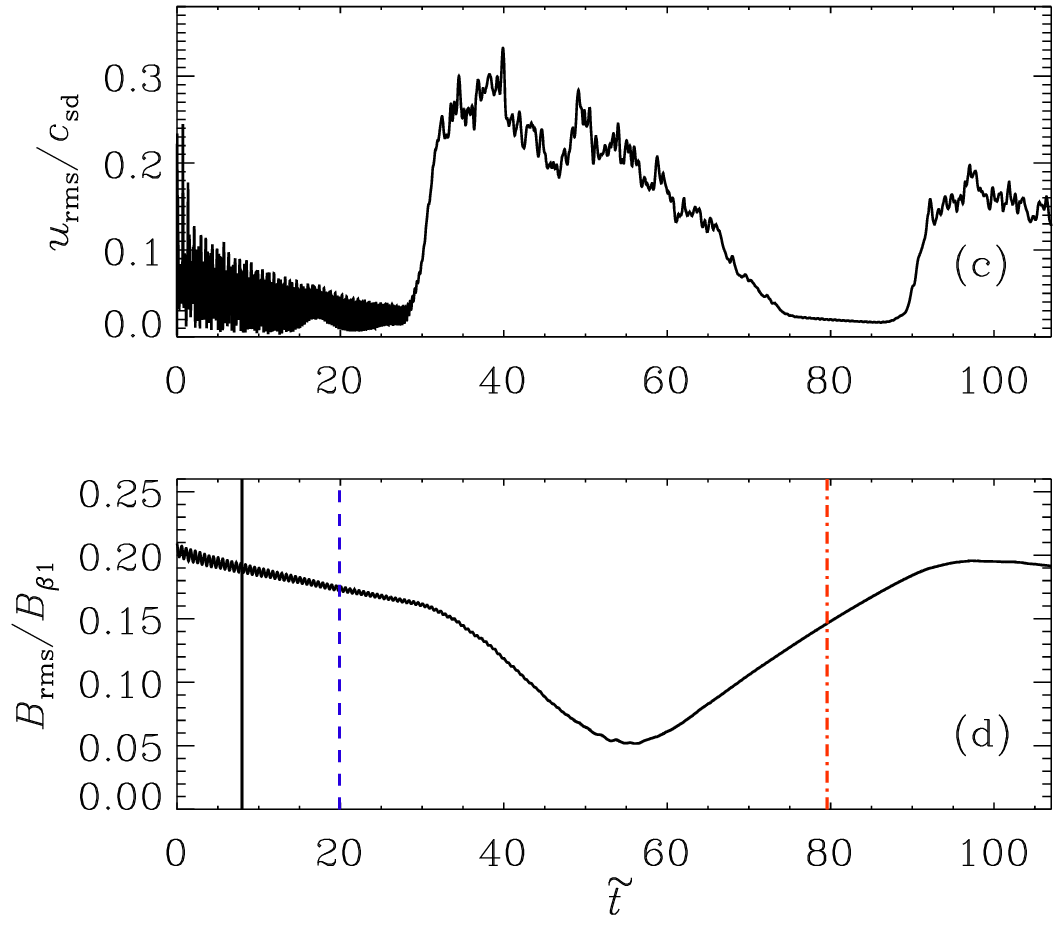}
\caption{Same as \Fig{profs1} but from the run A3e as listed in \Tab{tbl1}.
}
\label{ts_exp} \end{figure}

Here we show results from the run A3e listed in \Tab{tbl1},
where we adopt an explicit diffusion scheme
for the kinematic viscosity ($\nu$),
to demonstrate that the results presented in \Sec{res}
are robust. In \Fig{ts_exp}, we show the vertical profiles,
and the temporal evolution, of the thermodynamic variables.
Also shown are the time dependence of $\urms$ and $\Brms$.
The sharp transition between the cool disc and the hot corona
by the relaxation term in \Eq{equ:ss} is better maintained
in time in this case. Recall that, just like the other cases
discussed in this work, the run begins with a toroidal
field $B_y$ with a Gaussian profile along $\tilde{z}$ such that
the initial $\beta_0=5.7$ is constant within the disc.

\begin{figure}  
\centering
\includegraphics[width=0.48\columnwidth]{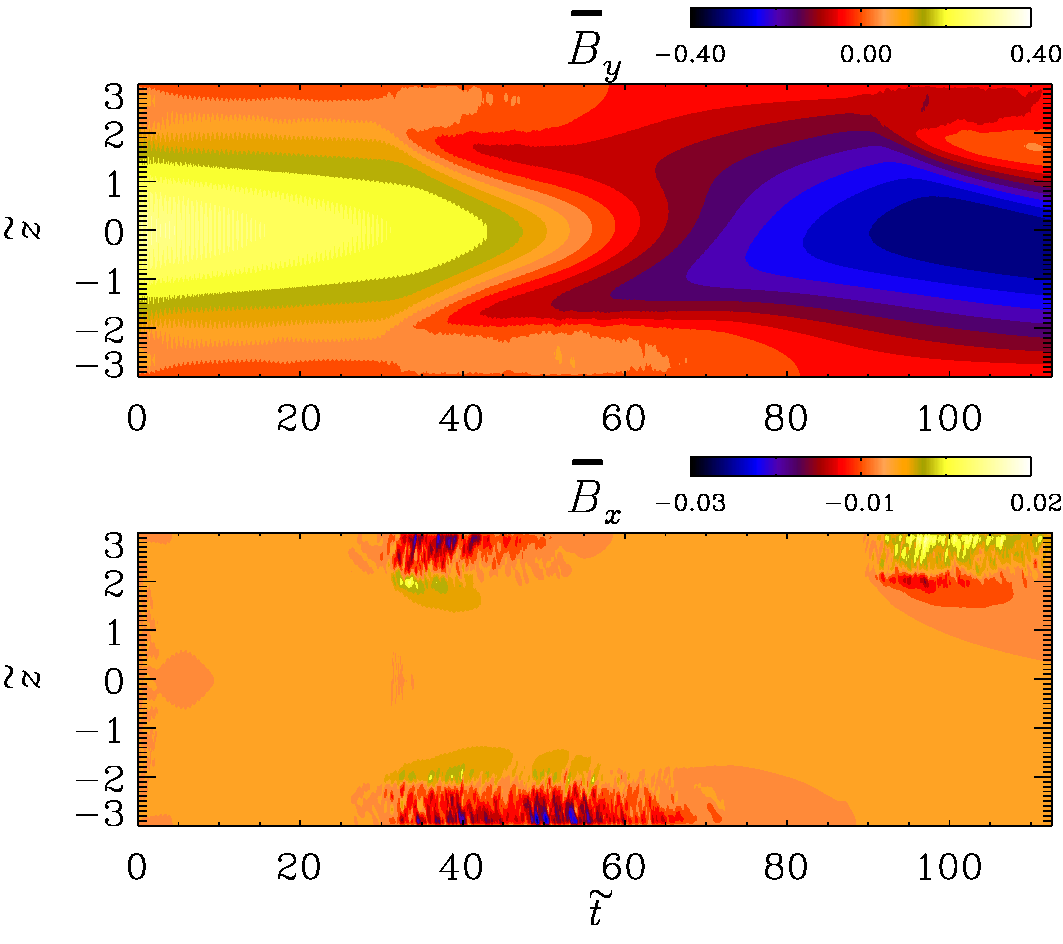}
\includegraphics[width=0.48\columnwidth]{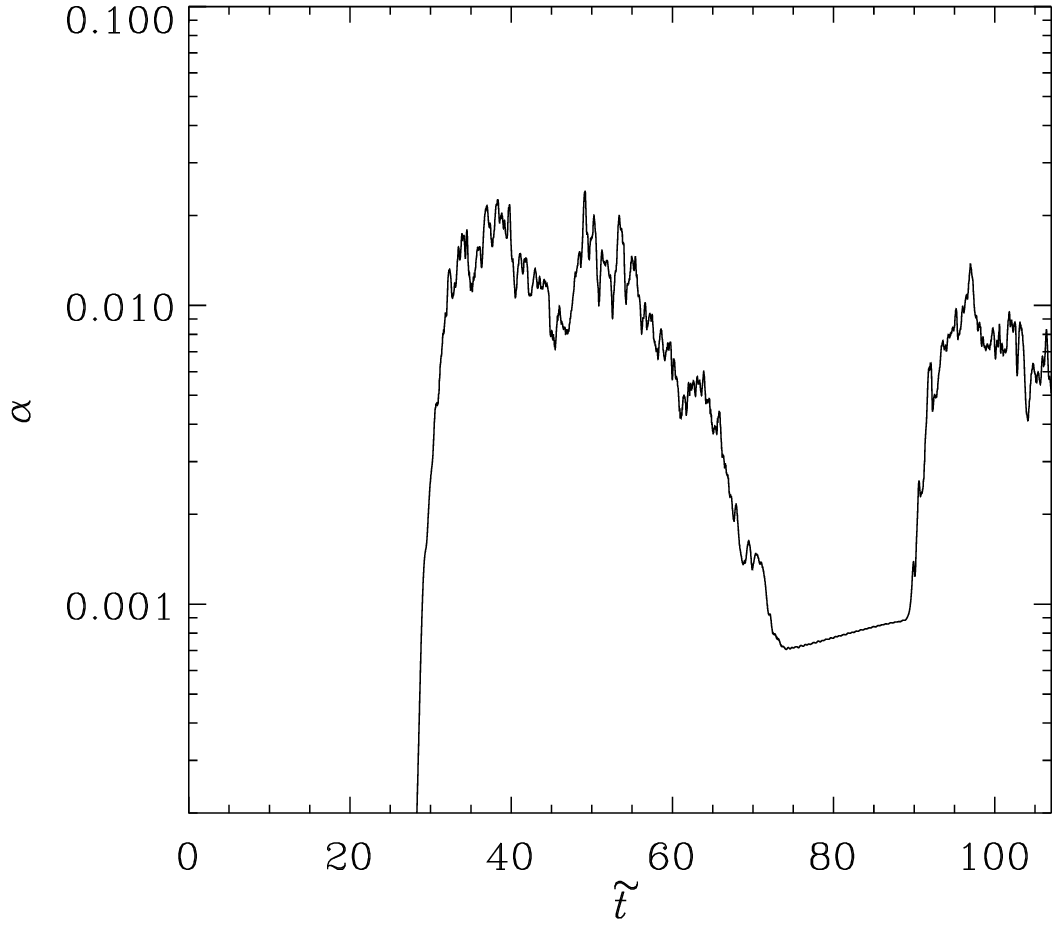}
\caption{
Same as \Fig{st_alp1} but from the run A3e as listed in \Tab{tbl1}.
}
\label{alpha_exp} \end{figure}

From \Fig{ts_exp}(c) and (d) we note that, while the
$\Brms$ evolves smoothly as in \Fig{profs1} for the
A1s$\chi$ model, $\urms$ shows an outburst like activity in
time. This leads to an interesting, burst-like temporal
evolution of the SS viscosity parameter $\alpha$ as shown
in the right panel of \Fig{alpha_exp}. Peak value of
the $\alpha$ exceeds the value 0.01 which is consistent with
our findings as presented in \Sec{res}, and also with
the values reported in some other works discussed before.

Spacetime diagrams of mean magnetic fields,
$\overline{B}_x$ and $\overline{B}_y$, as shown in
\Fig{alpha_exp} reveal the reversal of toroidal component
which is strongest in the disc region, whereas
$\overline{B}_x$ is strong in the coronal regions.
This is consistent with the results presented in
\Sec{res} where a number of cases with Smagorinsky scheme
for $\nu$ also show a similar pattern. Results presented
in this work are thus independent of the numerical
schemes for kinematic viscosity. Burst-like behavior of the
SS viscosity parameter $\alpha$ in time may have interesting
consequences for accretion pattern and may help us better
understand the observations. In future work, we will focus
more on this by performing simulations at larger Reynolds
numbers.

\end{document}